\documentclass{article}

\usepackage{arxiv}

\usepackage[utf8]{inputenc} 
\usepackage[T1]{fontenc}    
\usepackage{hyperref}       
\usepackage{url}            
\usepackage{booktabs}       
\usepackage{amsfonts,amsmath,amssymb}       
\usepackage{nicefrac}       
\usepackage{microtype}      
\usepackage{graphicx}
\usepackage{subcaption}
\usepackage{multirow}
\usepackage{multicol}

\title{Probabilistic elicitation of expert knowledge through assessment of computer simulations}

\author{
  Owen Thomas \\
  Department of Biostatistics\\
  University of Oslo\\
  Norway \\
  \texttt{o.m.t.thomas@medisin.uio.no} \\
    \And
  Henri Pesonen \\
  Department of Biostatistics\\
  University of Oslo\\
  Norway \\
  \texttt{henri.pesonen@medisin.uio.no} \\
   \And
  Jukka Corander \\
  Department of Biostatistics\\
  University of Oslo\\
  Norway \\
  \texttt{jukka.corander@medisin.uio.no} \\
}

\begin{document}
\maketitle

\begin{abstract}
We present a new method for probabilistic elicitation of expert knowledge using binary responses of human experts assessing simulated data from a statistical model, where the parameters are subject to uncertainty. The binary responses describe either the absolute realism of individual simulations or the relative realism of a pair of simulations in the two alternative versions of out approach. Each version provides a nonparametric representation of the expert belief distribution over the values of a model parameter, without demanding the assertion of any opinion on the parameter values themselves. Our framework also integrates the use of active learning to efficiently query the experts, with the possibility to additionally provide a useful misspecification diagnostic. We validate both methods on an automatic expert judging a binomial distribution, and on human experts judging the distribution of voters across political parties in the United States and Norway. Both methods provide flexible and meaningful representations of the human experts' beliefs, correctly identifying the higher dispersion of voters between parties in Norway.
\end{abstract}


\section{Introduction}

The challenge of accurately translating prior knowledge into prior probability distributions has loomed large in Bayesian statistics since its inception \cite{chesley1975elicitation,bernardo2009bayesian}. Methods exist to specify uninformative prior distributions in the absence of prior information, while empirical and objective Bayesian analysis aims to derive methods for prior specification from a data-driven or impersonal perspective \cite{berger2006case}.

\subsection{Prior Elicitation}

A subjective Bayesian analysis embraces the possibility of integrating expert opinion into the prior before data has been observed \cite{goldstein2006subjective}. Various methods have existed for eliciting prior information from experts, many of which focus on querying experts regarding different values of the statistical parameters \cite{o2006uncertain}. It is possible to ask the experts to make distributional assertions such as specifying the central tendency of the distribution, or regions of parameter space falling within given quantiles \cite{gosling2018shelf}. Many methods are concerned with selecting the hyperparameters of somewhat restrictive parametric representations of the experts' opinion, with some methods providing more complex belief distributions with hierarchical or nonparametric representations \cite{albert2012combining,oakley2007uncertainty}. Such methods have existed for some decades, and are used extensively within applied and industrial statistical projects \cite{johnson2010methods}.

The use of human feedback has been explored in a machine learning context under various names, including ``Interactive learning'' or ``human in the loop'' computation. The goal can be to improve predictive power of statistical models or enhance inference, or to build emulator models for human behaviour \cite{ware2001interactive,fails2003interactive,amershi2011effective, robert2016reasoning}. Such methods have been used in applications to inform the construction of statistical models using user preferences \cite{guo2010gaussian,ruotsalo2014interactive}.

A key concern when dealing with human interactions is to make effective use of their queries, owing to the finite resource of human time and energy. Several previous works have explored various methods for active learning of expert opinions conditional on previous responses, such that the expert responses are as informative as possible towards the elicitation problem \cite{afrabandpey2017interactive,daee2017knowledge,micallef2017interactive}.

\subsection{Classifiers and Likelihood-Free Inference}

Probabilistic classification has been established as a principled method for estimating a ratio of two distributions from which samples can be drawn, but are not necessarily analytically accessible \cite{sugiyama2012density}. A probabilistic classifier trained to discriminate between samples from each component in the ratio can be used to approximate the ratio by evaluating the odds ratio of the trained classifier on the data of interest. This has been used extensively in the context of sampling algorithms, but also recently receiving interest in the context of likelihood-free inference when classifying simulated data drawn from statistical models: such an approach has been used to evaluate ratios for likelihood proxies for neural networks \cite{goodfellow2014generative}, accept-reject ratios for sampling algorithms \cite{gutmann2018likelihood}, frequentist likelihood tests \cite{cranmer2015approximating}, likelihood-to-evidence Bayesian updates \cite{thomas2016likelihood,hermans2019likelihood}, and misspecification diagnosis\cite{thomas2019diagnosing}. Minimising the misclassification rate of the classifier has a principled interpretation as minimising an implicit statistical divergence between the two distributions defining the ratio \cite{sugiyama2012density2}.

The remainder of the paper is structured as follows. Section \ref{sec:meth} explains the methods developed in this paper, while Section \ref{sec:verify} presents the performance of the method on both automated and human expert data. The ultimate section concludes with a discussion and proposals of future work.

\section{PRECIOUS: PRior EliCItatiOn through jUdgements of Simulations}\label{sec:meth}

In this work, we extend the use of classifiers to extract the belief distribution of an expert, condition on binary judgements they provide on simulations drawn from the statistical model of interest. We propose two approaches: one in which experts are asked to judge whether simulations drawn from the model are credible draws from a real data set, and one in which pairs of simulations are compared with one another, implicitly targeting a different likelihood proxy.

\subsection{Verisimilitude Judgements - Veri-PRECIOUS}

When the expert is presented with simulations and asked to judge whether the data is a credible draw from reality, the resulting judgements can be used to train a Gaussian Process classifier that implicitly captures the expert's $\mathcal{E}$ belief distribution of parameters $\theta$  under the statistical model $\mathcal{M}$ that generate realistic simulations.

An expert $\mathcal{E}$ is shown a simulation $X_{\theta}\sim p(X|\theta,\mathcal{M})$ drawn from a model $\mathcal{M}$ given a parameter value $\theta$. The expert provides a binary data label $y_{\mathcal{E}}=1$ or $0$ depending on whether they think the simulation $X_{\theta}$ is realistic or not. Consequently, the set of independent and identically distributed labels $\mathbf{y}_\mathcal{E}$ are used to train a Gaussian Process classifier $\mathcal{C}$ to model $p(y|\theta)$, conditional on the model simulation parameters $\theta$:

\begin{align}
    p(y|f)=\Phi(f)\\
    p(f|\theta)\sim \mathcal{GP}(m(\theta),k(\theta,\theta'))\\
    p(y|\theta,\mathcal{C}) \sim \mathcal{GPC}(m(\theta),k(\theta,\theta'))
\end{align}

The data labels $y$ are connected to the latent function $f$ through a Bernoulli likelihood and a normal cdf link function $\Phi$. Inference over the latent variable $f$ is non-conjugate, so is now performed through Expectation Propagation, resulting in a tractable approximate Gaussian posterior.

We consider the expert $\mathcal{E}$ to be the base classifier in this instance, with the GP classifier acting as a proxy model for their decisions $\mathbf{y}_\mathcal{E}$. It has the effect of both providing probabilistic calibration to the binary responses of the expert and interpolating a smooth function between the responses of the expert at different values of $\theta$.

We consider the predictive probability of generating a realistic label $y=1$ as a function of the parameter $\theta$, conditional on the classifier $\mathcal{C}$, the information $\mathbf{y}_\mathcal{E}$ provided by the expert $\mathcal{E}$, and the model $\mathcal{M}$: $p(y|\theta,\mathcal{C},\mathbf{y}_\mathcal{E},\mathcal{M})$. It is possible to use Bayes' theorem to define a distribution $p(\theta|y=1,\mathcal{C},\mathbf{y}_\mathcal{E},\mathcal{M})$, indicating the distribution of $\theta$ that generate realistic simulated data:

\begin{align}
p(\theta|y=1,\mathcal{C},\mathbf{y}_\mathcal{E},\mathcal{M})=&\frac{p(y=1|\theta,\mathcal{C},\mathbf{y}_\mathcal{E},\mathcal{M})p(\theta|\mathcal{M})}{p(y=1|\mathcal{C},\mathbf{y}_\mathcal{E},\mathcal{M})}\\
=&\frac{p(y=1|\theta,\mathcal{C},\mathbf{y}_\mathcal{E},\mathcal{M})p(\theta|\mathcal{M})}{\int d\theta p(y=1|\theta,\mathcal{C},\mathbf{y}_\mathcal{E},\mathcal{M})p(\theta|\mathcal{M})}
\end{align}

We have introduced a distribution $p(\theta|\mathcal{M})$ that represents the belief distribution before integrating the information from the expert, and the marginal probability of the model $\mathcal{M}$ generating a realistic simulation $p(y=1|\mathcal{C},\mathbf{y}_\mathcal{E},\mathcal{M})$.

The marginal probability $p(y=1|\mathcal{C},\mathbf{y}_\mathcal{E},\mathcal{M})$ is a useful measure of the misspecification of the model $\mathcal{M}$ under the initial belief distribution $p(\theta|\mathcal{M})$, with a large value indicating a well-specified model that frequently generates simulations close to the true data, and a low value indicating the opposite.

We have consequently derived $p(\theta|y=1,\mathcal{C},\mathbf{y}_\mathcal{E},\mathcal{M})$, a nonparametric representation of the belief distribution over the parameter $\theta$, conditional on the classifier $\mathcal{C}$, the expert responses $\mathbf{y}_\mathcal{E}$, and the model $\mathcal{M}$, based only on their opinions on simulations $X_{\theta}$ and not directly on the parameter values themselves.

The latent Gaussian variable of the Gaussian Process classifier can also be used for active acquisition of the simulations to show the expert, making efficient use of the expert's judgements and energy. Many different acquisitions are possible: we use a standard Bayesian Optimisation Upper Confidence Bound acquisition function defined on the latent Gaussian variable, but others are also possible.

\subsection{Comparison Pairwise Judgements - Pari-PRECIOUS}

An alternative method is to present experts with pairs of simulations $[X_{\theta_1},X_{\theta_2}]$ drawn from different parameter values $\theta_1$ and $\theta_2$. The expert then provides a binary judgement label $y_\mathcal{E}$ indicating which simulation they consider more realistic: this lessens the relevance of model misspecification by only considering the relative merits of simulated data.

The expert's labels are again used to train a Gaussian Process Classifier with an additive kernel structure, separated for parameters $\theta_1$ and $\theta_2$:

\begin{equation}
    p(y|\theta_1, \theta_2,\mathcal{C}) \sim \mathcal{GPC}(m(\theta_1)+m(\theta_2),k(\theta_1,\theta_1'),k(\theta_2,\theta_2'))
\end{equation}

The corresponding odds ratio is an approximation of the likelihood ratio of the simulations under two different values of theta, which can be varied freely as predictive covariates of the classifier:

\begin{equation}\label{eqn:pari-PRECIOUS_p}
    \frac{p(y=1|\theta,\mathcal{C},\mathbf{y}_\mathcal{E})}{p(y=0|\theta,\mathcal{C},\mathbf{y}_\mathcal{E})}=  \frac{p(\theta_1|\mathcal{C},\mathbf{y}_\mathcal{E},\mathcal{M})}{p(\theta_2|\mathcal{C},\mathbf{y}_\mathcal{E},\mathcal{M})}
\end{equation}

In practice, the belief distribution in Equation \eqref{eqn:pari-PRECIOUS_p} is evaluated with the likelihood-maximising value of $\theta$ used as $\theta_2$ in the denominator for stability reasons. 

The acquisition space now spans the two versions of parameter space $\theta_1$ and $\theta_2$, leading to a more complex acquisition procedure. There is a clear symmetry to the space, so acquisitions can be limited to $\theta_1<\theta_2$ without loss of generality, but a standard Bayesian Optimisation procedure will not be appropriate. Instead, we use Preferential Bayesian Optimisation, which provides pairs of points for informative comparison: the most likely value of $\theta$ is chosen for one of the simulations, with the other chosen as the value with most associated uncertainty, conditional on the first value \cite{gonzalez2017preferential}.

\section{Examples and Results}\label{sec:verify}

\subsection{Binomial Distribution with an Automated Expert}

In this section we consider a simple example of data sampled from a binomial model $\mathcal{B}(n,q)$ with $n=100$ trials and success probability $q$, interpreted as 100 tosses of a coin with an unknown bias. Draws from the binomial distribution conditioned on $q$ are then shown to an expert with an opinion on the value of $q$. The expert will then accept or reject the simulated coin tosses according to whether they coincide with their belief on $q$.

This example can also be performed with an ``automated expert'' to facilitate demonstration of the elicitation procedure. For the purposes of this example, the expert assumes a true value of $q=0.5$ and for the realism judgements, samples with fewer than 35 or more than 65 heads are considered unrealistic and are rejected. For the preference comparisons then the simulation with number of heads closest to 50 is preferred. A uniform prior $p(q|\mathcal{M})$ was assumed.

We ran a simulated belief elicitation procedure with the automated experts. For the realism judgements, we used an initial grid of 21 points evenly spaced to initialise the Bayesian Optimisation procedure, and a further 79 points drawn with informative acquisitions according to the procedure targeting regions of high likelihood-proxy variance described earlier.

For the preference judgements, we used an initial half grid of 15 points to initialise the algorithm over the two-dimensional preference space defined over the one-dimensional parameter $q$, with a further 85 points drawn according to the preference acquisition function, bringing the total number of expert judgements to 100.

\begin{figure}
    \centering
    \begin{subfigure}[b]{0.45\textwidth}
        \includegraphics[width=\textwidth]{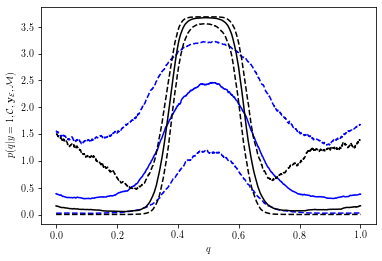}
        \caption{$p(q|y=1,\mathcal{C},\mathbf{y}_\mathcal{E},\mathcal{M})$ from Veri-PRECIOUS}
        \label{fig:veri-binomial}
    \end{subfigure}
    ~ 
    \begin{subfigure}[b]{0.45\textwidth}
        \includegraphics[width=\textwidth]{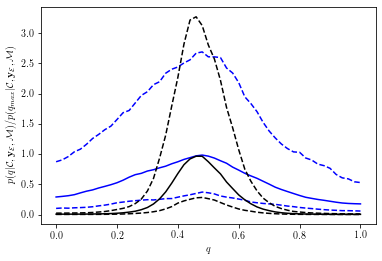}
        \caption{$p(q|\mathcal{C},\mathbf{y}_\mathcal{E},\mathcal{M})/p(q_\text{max}|\mathcal{C},\mathbf{y}_\mathcal{E},\mathcal{M})$ from Pari-PRECIOUS}
        \label{fig:pari-binomial}
    \end{subfigure}
    \caption{Belief distributions with pointwise 10\% and 90\% quantiles, derived from an automatic expert assuming $q=0.5$, judging simulations from a binomial distribution with 100 draws. Distributions after the initial grid of acquisitions are shown in blue, and those after active acquisitions are shown in black.}\label{fig:binomial_distributions}
\end{figure}

The final likelihood proxies derived from the expert responses are presented in Figure \ref{fig:binomial_distributions}, alongside the belief distributions after the initial grid of acquisitions but before the active learning. The preference proxy has a denominator defined by the maximum of the likelihood proxy, i.e. $p(q|y)/p(q_\text{max}|y)$. We see that both Veri-PRECIOUS and Pari-PRECIOUS provide similar but distinct likelihood proxies, with Veri-PRECIOUS having a somewhat more distinct shoulders than that generated from Pari-PRECIOUS. Both successfully characterise the mean and mode of the expert belief of an unbiased coin with $q=0.5$, and a reasonable distributions of uncertainty around the same central tendency. The veri-PRECIOUS belief distribution exhibits small amounts of growth in the tails where the uncertainty increases away from the acquired data. The quantiles presented show that this is clearly interpretable as regression to the GP prior distribution, and veri-PRECIOUS is not making strong claims as to the functional form of the belief in the tails. Such effects can be minimised through the use of a prior distribution with any decrease in the tails.

\begin{figure}
    \centering
    \begin{subfigure}[b]{0.45\textwidth}
        \includegraphics[width=\textwidth]{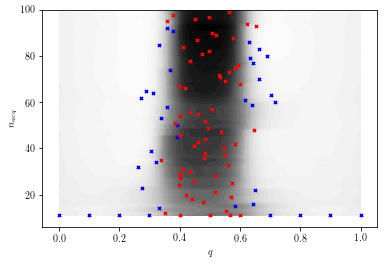}
        \caption{A contour map of $p(q|y=1,\mathcal{C},\mathbf{y}_\mathcal{E},\mathcal{M})$ from Veri-PRECIOUS changing with $n_\text{acq}$}
        \label{fig:veri-binomial_dynamic_lik}
    \end{subfigure}
    ~ 
    \begin{subfigure}[b]{0.45\textwidth}
        \includegraphics[width=\textwidth]{binomial_veri_dynamic_lik.png}
        \caption{A contour map of the acquisition function for Veri-PRECIOUS changing with $n_\text{acq}$}
        \label{fig:veri-binomial_dynamic_acq}
    \end{subfigure}
    \caption{Veri-Precious belief distribution and acquisition function changing with $n_\text{acq}$, with an automatic expert assuming $q=0.5$, judging simulations from $\mathcal{B}(100,q)$, red being labelled realistic and blue unrealistic.}\label{fig:binomial_distributions_dynamic_veri}
\end{figure}

\begin{figure}
    \centering
    \begin{subfigure}[b]{0.45\textwidth}
        \includegraphics[width=\textwidth]{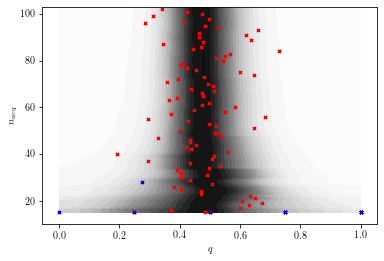}
        \caption{A contour map of $p(q|\mathcal{C},\mathbf{y}_\mathcal{E},\mathcal{M})/p(q_\text{max}|\mathcal{C},\mathbf{y}_\mathcal{E},\mathcal{M})$ from Pari-PRECIOUS changing with $n_\text{acq}$}
        \label{fig:pari-binomial_dynamic_lik}
    \end{subfigure}
    ~ 
    \begin{subfigure}[b]{0.45\textwidth}
        \includegraphics[width=\textwidth]{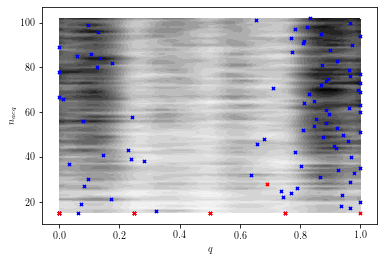}
        \caption{A contour map of the conditional uncertainty acquisition function for Pari-PRECIOUS changing with $n_\text{acq}$}
        \label{fig:pari-binomial_dynamic_acq}
    \end{subfigure}
    \caption{Veri-Precious belief distribution and acquisitions function changing with $n_\text{acq}$, with an automatic expert assuming $q=0.5$, judging simulations from $\mathcal{B}(100,q)$. The belief distribution is used as an acquisition function for the first member of the preference comparison, and the uncertainty conditional on the first is used as an acquisition for the second. Acquisitions plotted in red were judged preferable in the comparison, blue were less realistic.}\label{fig:binomial_distributions_dynamic_pari}
\end{figure}

The belief distributions and acquisition functions are represented with increasing number of acquisitions $n_\text{acq}$ in Figures \ref{fig:binomial_distributions_dynamic_veri} and \ref{fig:binomial_distributions_dynamic_pari}. The belief distributions can be seen to converge to sharper distributions as greater numbers of expert labels are extracted in Figures \ref{fig:veri-binomial_dynamic_lik} and \ref{fig:pari-binomial_dynamic_lik}. The acquisitions can similarly be seen to stabilise for Veri-PRECIOUS in Figure \ref{fig:veri-binomial_dynamic_acq}, with the acquisitions process converging to realistic simulations plotted in red drawn from near the belief distribution mode.

The more complex acquisition process for Pari-PRECIOUS is illustrated in Figure \ref{fig:binomial_distributions_dynamic_pari}, with the maximum of the belief distribution shown in Figure \ref{fig:pari-binomial_dynamic_lik} being used to select the first parameter in the comparison, and the variance conditional on the first parameter value shown in Figure \ref{fig:pari-binomial_dynamic_acq} used to select the second. The acquired values are also plotted, generally consisting of one parameter value near the belief distribution mode and one towards the tails. The preferred parameter values are plotted in red, acquired mostly from the mode of the belief distribution.

\subsection{Political Voter Affiliation Perception of Norwegian and American Politics}

\begin{figure}
    \centering
    \includegraphics{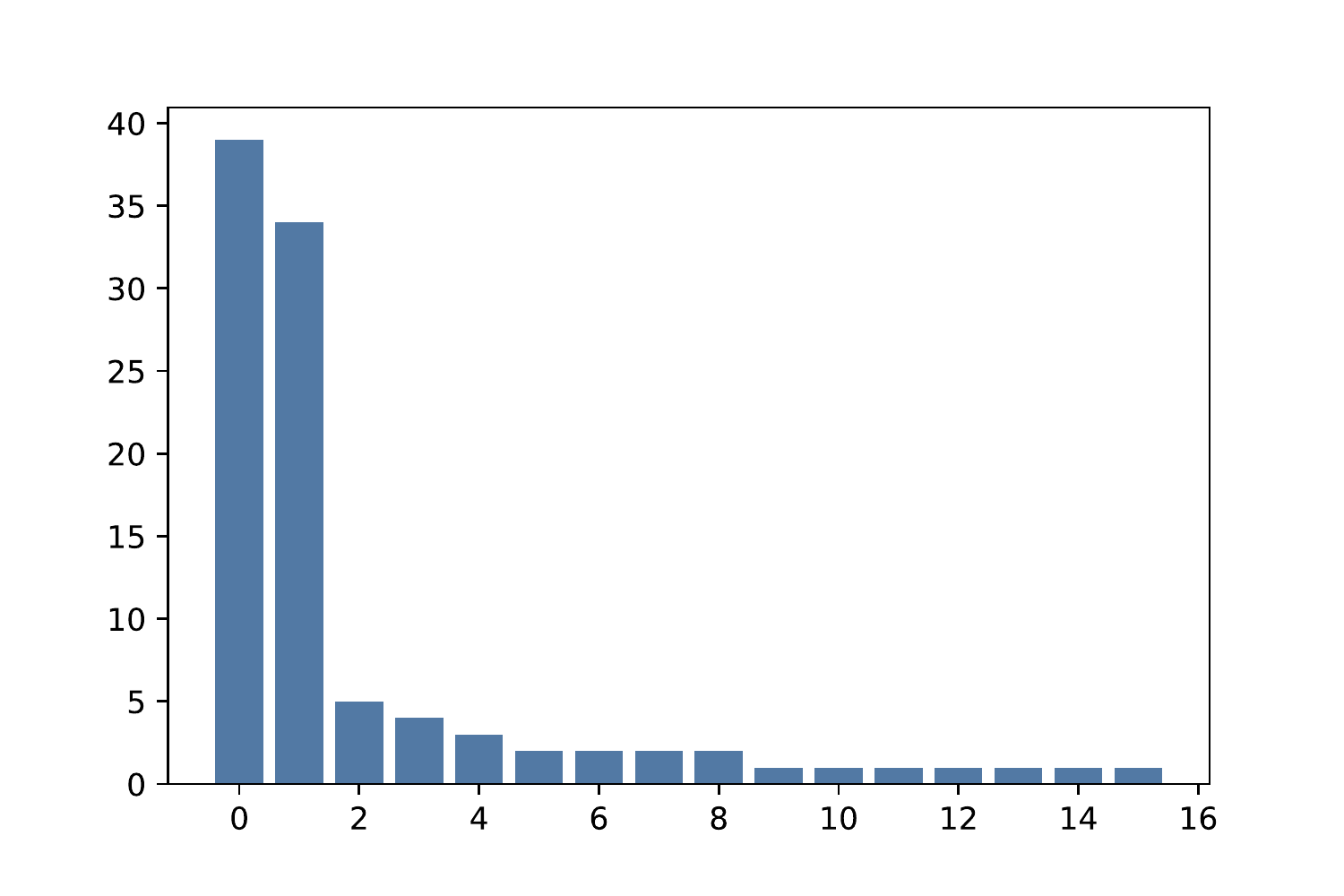}
    \caption{An example simulation of 100 individuals assigned into groups, generated from a Dirichlet Process with an $\alpha$ dispersion hyperparameter value equal to 0.5, plotted in the same way as was presented to the experts. This would be interpreted that fifteen different parties were represented among the 100 respondents, with a large majority voting for the two main parties.}
    \label{fig:CRP_example}
\end{figure}

Here we consider human experts and their opinions on the distribution of voters according to the political parties they voted for in the most recent election in their countries. We use a Dirichlet Process to model the association of individuals to political parties, with a sample of 100 individuals assigned to groups, conditional on a single dispersion hyperparameter $\alpha$ defined between zero and one \cite{teh2010dirichlet}. Larger values of $\alpha$ correspond to more heavily dispersed assignment of individuals to clusters. One draw of 100 individuals for $\alpha=0.5$ is shown in Figure \ref{fig:CRP_example}. A uniform distribution $p(\alpha|\mathcal{M})\sim\mathcal{U}(0,1)$ is used as the belief representation before the information from the experts.

The experts are then asked to judge whether the simulations accurately represent party membership of 100 members of the voting public drawn at random from a random neighbourhood in their country, and were reminded of the noise introduced by sampling a relatively small number of people, as well as the potential heterogeneity of voters across neighbourhoods. We recruited five Norwegian people who were separately asked to compare the simulations with their perceptions of politics in Norway and the United States of America. 

Norway has a more diverse multiparty democratic system than the USA: in the most recent elections, 9 Norwegian political parties gained more than 1\% of the popular vote \cite{Norwayelection}, compared to 4 candidates in the most recent presidential election in the USA \cite{USAelection}. Similarly, the seven most popular parties in Norway each gained more than 4\% of the vote, whereas only two parties' candidates in the USA achieved the same. Consequently, we would expect a higher value of the dispersion parameter $\alpha$ to be more appropriate to describe the distribution of voters across parties in Norway compared to the voters in the USA.

Participants were queried through both the veri-PRECIOUS framework and the pari-PRECIOUS framework, considering about the verisimilitude of individual simulations and the relative realism of pairs of simulations, respectively. A two-stage acquisition process used, with an initial grid of parameter values being used, followed by an active learning procedure using a Bayesian Optimisation process to effectively query parameter values. Experts were shown a total of 50 simulations for veri-PRECIOUS and 50 pairs of simulations for pari-PRECIOUS.

\begin{figure}
    \centering
    \begin{subfigure}[b]{0.4\textwidth}
        \includegraphics[width=\textwidth]{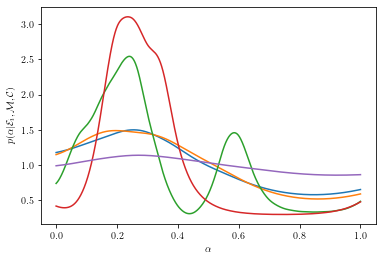}
        \caption{Individual USA $p(\alpha|y=1,\mathbf{y}_{\mathcal{E}_i},\mathcal{M},\mathcal{C})$ from Veri-PRECIOUS}
        \label{fig:indie-usa-veri}
    \end{subfigure}
    ~ 
    \begin{subfigure}[b]{0.4\textwidth}
        \includegraphics[width=\textwidth]{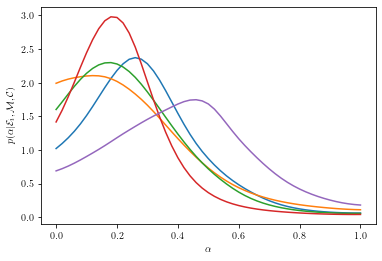}
        \caption{Individual USA $p(\alpha|\mathbf{y}_{\mathcal{E}_i},\mathcal{M},\mathcal{C})$ from Pari-PRECIOUS}
        \label{fig:indie-usa-pari}
    \end{subfigure}
    \begin{subfigure}[b]{0.4\textwidth}
        \includegraphics[width=\textwidth]{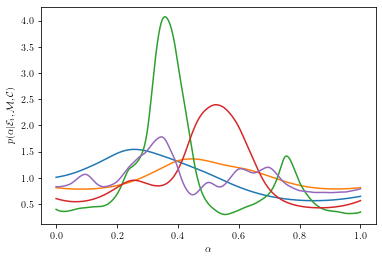}
        \caption{Individual Norway $p(\alpha|y=1,\mathbf{y}_{\mathcal{E}_i},\mathcal{M},\mathcal{C})$ from Veri-PRECIOUS}
        \label{fig:indie-norway-veri}
    \end{subfigure}
    ~ 
    \begin{subfigure}[b]{0.4\textwidth}
        \includegraphics[width=\textwidth]{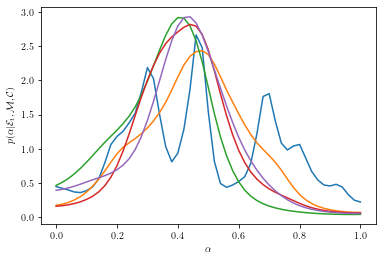}
        \caption{Individual Norway $p(\alpha|\mathbf{y}_{\mathcal{E}_i},\mathcal{M},\mathcal{C})$ from Pari-PRECIOUS}
        \label{fig:indie-norway-pari}
    \end{subfigure}
    \begin{subfigure}[b]{0.4\textwidth}
        \includegraphics[width=\textwidth]{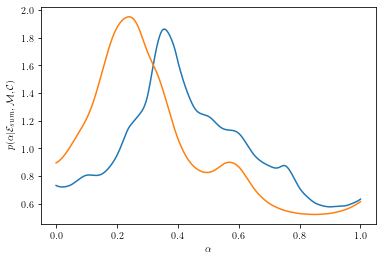}
        \caption{Averaged USA and Norway $p(\alpha|\mathcal{E}_\text{sum},\mathcal{M},\mathcal{C})$ from Veri-PRECIOUS in orange and blue, respectively.}
        \label{fig:soe_average_political_veri}
    \end{subfigure}
    ~ 
    \begin{subfigure}[b]{0.4\textwidth}
        \includegraphics[width=\textwidth]{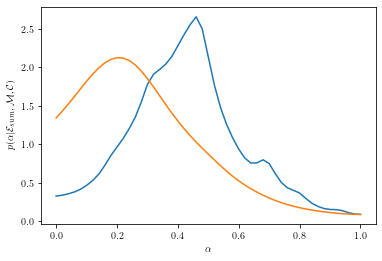}
        \caption{Averaged USA and Norway $p(\alpha|\mathcal{E}_\text{sum},\mathcal{M},\mathcal{C})$ from Pari-PRECIOUS in orange and blue, respectively.}
        \label{fig:soe_average_political_pari}
    \end{subfigure}
    \begin{subfigure}[b]{0.4\textwidth}
        \includegraphics[width=\textwidth]{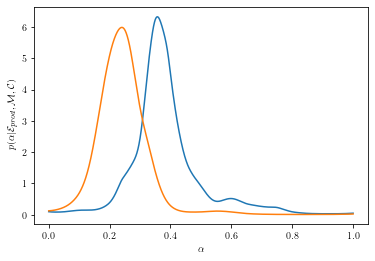}
        \caption{Averaged USA and Norway $p(\alpha|\mathcal{E}_\text{prod},\mathcal{M},\mathcal{C})$ from Veri-PRECIOUS in orange and blue, respectively.}
        \label{fig:poe_average_political_veri}
    \end{subfigure}
    ~ 
    \begin{subfigure}[b]{0.4\textwidth}
        \includegraphics[width=\textwidth]{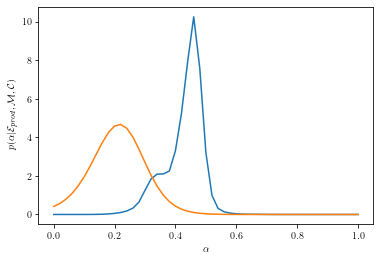}
        \caption{Averaged USA and Norway $p(\alpha|\mathcal{E}_\text{prod},\mathcal{M},\mathcal{C})$ from Pari-PRECIOUS in orange and blue, respectively.}
        \label{fig:poe_average_political_pari}
    \end{subfigure}
\caption{Individual $p(\alpha|\mathbf{y}_{\mathcal{E}_i},\mathcal{M},\mathcal{C})$ and averaged $p(\alpha|\mathcal{E}_\text{sum},\mathcal{M},\mathcal{C})$ and $p(\alpha|\mathcal{E}_\text{prod},\mathcal{M},\mathcal{C})$ belief distributions elicited by veri-PRECIOUS and pari-PRECIOUS when querying experts about the distribution of voters in the USA and Norway.}        \label{fig:political_beliefs}
\end{figure}

The resulting distributions elicited from the experts are plotted for the veri-PRECIOUS and pari-PRECIOUS in Figure \ref{fig:political_beliefs}. The sum-of-experts unweighted averages of the belief distributions concerning Norway and the USA can be used to represent the joint belief, defined by a pointwise average over the distributions of the individual experts $\mathcal{E}_i$ at each value of $\alpha$:
\begin{equation}\label{eqn:soe_average_belief}
    p(\alpha|\mathcal{E}_\text{sum},\mathcal{M},\mathcal{C})\propto\sum_i p(\alpha|\mathbf{y}_{\mathcal{E}_i},\mathcal{M},\mathcal{C})
\end{equation}

 $p(\alpha|\mathcal{E}_\text{sum},\mathcal{M},\mathcal{C})$ for the respondents are presented in Figures \ref{fig:soe_average_political_veri} and \ref{fig:soe_average_political_pari}. Average distributional means $\mu_\text{sum}$ and standard deviations $\sigma_\text{sum}$ from the averaged distributions $p(\alpha|\mathcal{E}_\text{sum},\mathcal{M},\mathcal{C})$ are presented in Table \ref{tab:political_results}.

We see in Figures \ref{fig:soe_average_political_veri} and \ref{fig:soe_average_political_pari} that the sum-of-experts averaged belief distributions concerning Norwegian politics put more probability mass on larger values of $\alpha$ than the averaged belief distributions concerning politics in the USA, suggesting that both methods have successfully elicited an meaningful difference in beliefs between the two sets of experts.

We notice in Figures \ref{fig:indie-usa-veri} and \ref{fig:indie-norway-veri} that some of the respondents were quite skeptical in their responses to veri-PRECIOUS, describing all but a few of the simulations as unrealistic, resulting in quite flat elicited belief distributions, whereas all of the belief distributions elicited by pari-PRECIOUS are quite informative.

The additional flatter elicited belief from the respondents in veri-PRECIOUS may have influenced the averaged belief distribution towards the uniform prior belief $p(\alpha|\mathcal{M})$. In order to investigate this effect, we used an additional method to combine the individual belief distributions, taking a product-of-experts combination of the expert's belief distribution:

\begin{equation}\label{eqn:poe_average_belief}
    p(\alpha|\mathcal{E}_\text{prod},\mathcal{M},\mathcal{C})\propto\prod_i p(\alpha|\mathbf{y}_{\mathcal{E}_i},\mathcal{M},\mathcal{C})
\end{equation}

The product-of-experts distribution $p(\alpha|\mathcal{E}_\text{prod},\mathcal{M},\mathcal{C})$ is more heavily influenced by the more informative individual experts, reducing the influence of uniform beliefs. The product-of-experts distributions are presented in Figures \ref{fig:poe_average_political_veri} and \ref{fig:poe_average_political_pari}. Means and standard deviations $\mu_\text{prod}$ and $\sigma_\text{prod}$ derived from the product-of-experts distribution presented in Table \ref{tab:political_results}. 

\begin{table}[]
    \centering
    \begin{tabular}{cc|cc|cc}
         && \multicolumn{2}{c}{USA}&\multicolumn{2}{c}{Norway}  \\
         &&mean&s.d.&mean&s.d.\\
\multirow{2}{*}{veri-PRECIOUS}&$\mu_\text{sum},\sigma_\text{sum}$&0.399&0.262&0.464&0.252\\
&$\mu_\text{prod},\sigma_\text{prod}$&0.243& 0.0890 &0.387&0.121\\
\multirow{2}{*}{pari-PRECIOUS}&$\mu_\text{sum},\sigma_\text{sum}$&0.291&0.207&0.419&0.196\\
&$\mu_\text{prod},\sigma_\text{prod}$&0.208&0.0916&0.421&0.0847\\
    \end{tabular}
    \caption{Derived means and standard deviation for the combined elicited beliefs, using $\mu_\text{sum},\sigma_\text{sum}$ from the distribution defined in Equation \ref{eqn:soe_average_belief}, and $\mu_\text{prod},\sigma_\text{prod}$ from that defined in Equation \ref{eqn:poe_average_belief}. }
    \label{tab:political_results}
\end{table}

We see in Table \ref{tab:political_results} that there is also a noticeable difference between the USA and Norway means when the influence of the more skeptical experts is downweighted. We also observe in Figures \ref{fig:poe_average_political_veri} and \ref{fig:soe_average_political_pari} that there are noticeable differences in the aggregated belief distributions of the USA and Norway beliefs when using pari-PRECIOUS, for which model misspecification and expert skepticism is less influential. Every combination of methods for elicitation and opinion combination returns a smaller aggregated estimate of $\alpha$ for opinions concerning USA compared to Norway, suggesting that a consistent and empirically expected result is being extracted.

\section{Discussion}

In this work, we have successfully demonstrated the use of Gaussian Process classifiers to elicit belief distributions from experts providing binary feedback. We have used active learning techniques to make efficient acquisition of expert opinions. The use of binary labels is intuitive for a human user, and can be effectively combined with a Gaussian Process Classifier to generated non-parametric representations of the expert's internal belief distribution. We demonstrate how to successfully use binary human judgements of whether individual simulations are realistic in veri-PRECIOUS, and also a pairwise comparison of the relative merits of two simulations in pari-PRECIOUS. Each of veri-PRECIOUS or pari-PRECIOUS may be more appropriate for a given application. We also note that veri-PRECIOUS provides a useful indicator of model misspecification with the marginal probability of generating a realistic simulation $p(y=1|\mathcal{C},\mathcal{E},\mathcal{M})$.

We validated our methodology on an ``automatic expert'' judging simulations from a binomial model, generating consistent and sensible distributional representations of the automatic experts implicit belief.

We further validated our methodology on human participants for the estimation of the otherwise hard-to-interpret dispersion parameter $\alpha$ of a Dirichlet Process for a real-world political example. The extracted belief distributions return a consistently lower estimate of the voter dispersion $\alpha$ for beliefs elicited concerning the USA compared to Norway, which is consistent with known properties of the multi-party system in each country.

\section{Acknowledgements}

We thank Kusti Skyt\'{e}n, Teemu Kallonen, Juri Kuronen, Jan Kokko, Rebecca Gladstone, David Kirk, Harry Thorpe, John Lees, Umberto Simola, Morten Valberg, Leiv R{\o}nneberg, P{\aa}l Christie Ryalen, Emilie {\O}degaard and Mari Brathovde for acting as experts for various iterations of the algorithm. We thank Gerry Tonkin-Hill for constructive discussions of the project, Waldir Leoncio for help with computational implementation, and David Balding for acting as a host at the University of Melbourne where the methodology was conceived. The authors gratefully acknowledge the support of the European Research Council [742158].

\bibliographystyle{unsrt}  
\bibliography{template}  


\end{document}